\pdfoutput=1
\documentclass[10pt, a4paper]{article}
\usepackage{lrec2022} 
\usepackage{multibib}
\newcites{languageresource}{Language Resources}
\usepackage{graphicx}
\usepackage[table,xcdraw]{xcolor}
\newcolumntype{?}[1]{!{\vrule width #1}}
\usepackage{tabularx}
\usepackage{multirow}
\usepackage{booktabs}
\usepackage{makecell}
\usepackage{tipa}
\usepackage[T1]{fontenc}
\usepackage{soul}
\usepackage{titlesec}
\titleformat{\section}{\normalfont\large\bfseries\center}{\thesection.}{1em}{}
\titleformat{\subsection}{\normalfont\SmallTitleFont\bfseries\raggedright}{\thesubsection.}{1em}{}
\titleformat{\subsubsection}{\normalfont\normalsize\bfseries\raggedright}{\thesubsubsection.}{1em}{}
\renewcommand\thesection{\arabic{section}}
\renewcommand\thesubsection{\thesection.\arabic{subsection}}
\renewcommand\thesubsubsection{\thesubsection.\arabic{subsubsection}}

\usepackage{epstopdf}
\usepackage[utf8]{inputenc}
\usepackage{newtxmath,newtxtext}

\usepackage{hyperref}
\usepackage{xstring}

\usepackage{color}

\title{Common Phone: A Multilingual Dataset for Robust Acoustic Modelling}

\name{Philipp Klumpp\textsuperscript{1}, Tom\'{a}s Arias-Vergara\textsuperscript{1,2}, Paula-Andrea P\'{e}rez-Toro\textsuperscript{1,2}\\{\bf \large Elmar N\"{o}th\textsuperscript{1}, Juan Rafael Orozco-Arroyave\textsuperscript{1,2}}} 

\address{\textsuperscript{1}Friedrich-Alexander-Universit\"{a}t Erlangen-N\"{u}rnberg, \textsuperscript{2}Universidad de Antioquia\\
         \textsuperscript{1}Lehrstuhl f\"{u}r Mustererkennung, Martensstrasse 3, 91058 Erlangen, Germany\\
         \textsuperscript{2}Calle 67 \# 53-108, UdeA Campus Principal, Medell\'{i}n, Colombia\\
         philipp.klumpp@fau.de\\}

\abstract{
Current state of the art acoustic models can easily comprise more than 100 million parameters. This growing complexity demands larger training datasets to maintain a decent generalization of the final decision function. An ideal dataset is not necessarily large in size, but large with respect to the amount of unique speakers, utilized hardware and varying recording conditions. This enables a machine learning model to explore as much of the domain-specific input space as possible during parameter estimation. This work introduces \textit{Common Phone}, a gender-balanced, multilingual corpus recorded from more than 11,000 contributors of \textit{Mozilla's Common Voice} project. It comprises around 116 hours of speech enriched with automatically generated phonetic segmentation. A Wav2Vec 2.0 acoustic model was trained with the \textit{Common Phone} to perform phonetic symbol recognition and validate the quality of the generated phonetic annotation. The architecture achieved a PER of 18.1\,\% on the entire test set, computed with all 101 unique phonetic symbols, showing slight differences between the individual languages. We conclude that \textit{Common Phone} provides sufficient variability and reliable phonetic annotation to help bridging the gap between research and application of acoustic models.
 \\ \newline  }

\begin{document}

\maketitleabstract

\section{Introduction}
In the past two years, Wav2Vec~\cite{schneider2019wav2vec} and Wav2Vec 2.0~\cite{baevski2020wav2vec} have leveraged the state-of-the-art of acoustic models to a new level, with the latter being able to achieve a Phoneme Error Rate (PER) of 8.3\,\% on the TIMIT~\citelanguageresource{garofolo1993darpa} test set. TIMIT is among the most important corpora for acoustic model evaluation, specifically because of the precise, manually annotated, phonetic reference. All speakers had been recorded in equal acoustic conditions using the same microphone. For acoustic model validation, these conditions can be considered perfect. However, the constricted acoustic environment also implies limited robustness to altered recording conditions. The corpus is also not gender-balanced, with 439 (~70\,\%) of the 630 speakers being male. For any application-driven project, the total number of speakers is quite small as well for modern standards.\\
A large variability of training samples is key to enable a deep architecture to explore as much of the input space as possible. Any perturbation in said space could result in the model performance to collapse~\cite{wang2018towards}. Deviations from the \textit{known input-space} could be so small that it would be impossible for a human rater to perceive them, thus they might be exploited in scenarios like adversarial attacks~\cite{schonherr2018adversarial,hu2019adversarial}. Robustness could be improved during training through techniques such as weight regularization, drop-out or batch-normalization~\cite{kukavcka2017regularization}. Another way would be to employ a dataset that provides a large variability with respect to acoustic conditions, recording hardware, contributing speakers, dialects and other parameters.\\
One such corpus could be \textit{Common Voice}~\citelanguageresource{ardila2019common} (CV). It is an ongoing initiative maintained by \textit{Mozilla Foundation} that aims to collect spoken text samples from contributors of many different languages. Everyone could donate their speech to enrich the corpus via the project website\,\footnote{\url{https://commonvoice.mozilla.org/}}. The most recent release \textit{7.0} from July 2021 comprised datasets in 76 different languages. Contributions could be made anonymously, additional information such as age and gender could be provided after registration on the website. Volunteers are not limited to donate their speech to help improve the quality of the corpus. They could also help to validate new speech donations, e. g. verify that the spoken text in an audio sample matches with the prompted text transcript. This crowd-based approach of automated donation and validation procedures enabled CV to collect data from a large amount of speakers. For example, the English \textit{7.0} corpus comprises more than 75,000 different speakers.

\newcolumntype{b}{X}
\newcolumntype{s}{X}

\begin{table*}[t]
\centering
\footnotesize
\rowcolors{2}{white}{gray!15}
\begin{tabularx}{1.0\textwidth}{b?{1.0pt}s|s|s|s|s|s|s|s}
\toprule
\rowcolor{white}
 & \makecell{Age~\textless~19} & \makecell{19~-~29} & \makecell{30~-~39} & \makecell{40~-~49} & \makecell{50~-~59} & \makecell{60~-~69} & \makecell{70~-~79} & \makecell{80~-~89} \\
\hline
English  & \makecell{276 \\ 276} & \makecell{867 \\ 870} & \makecell{417 \\ 421} & \makecell{270 \\ 274} & \makecell{287 \\ 290} & \makecell{170 \\ 173} & \makecell{57 \\ 58} & \makecell{5 \\ 5} \\
German  & \makecell{24 \\ 24} & \makecell{127 \\ 127} & \makecell{94 \\ 95} & \makecell{46 \\ 50} & \makecell{71 \\ 72} & \makecell{28 \\ 28} & \makecell{5 \\ 5} & \makecell{0 \\ 0} \\
Spanish  & \makecell{72 \\ 72} & \makecell{263 \\ 263} & \makecell{110 \\ 110} & \makecell{76 \\ 76} & \makecell{55 \\ 55} & \makecell{12 \\ 12} & \makecell{0 \\ 0} & \makecell{0 \\ 0} \\
French  & \makecell{38 \\ 38} & \makecell{186 \\ 186} & \makecell{90 \\ 90} & \makecell{78 \\ 78} & \makecell{76 \\ 77} & \makecell{41 \\ 41} & \makecell{6 \\ 6} & \makecell{0 \\ 0} \\
Italian  & \makecell{9 \\ 9} & \makecell{82 \\ 82} & \makecell{53 \\ 53} & \makecell{42 \\ 42} & \makecell{42 \\ 42} & \makecell{22 \\ 22} & \makecell{4 \\ 4} & \makecell{0 \\ 0} \\
Russian  & \makecell{18 \\ 18} & \makecell{48 \\ 48} & \makecell{17 \\ 17} & \makecell{9 \\ 9} & \makecell{3 \\ 3} & \makecell{0 \\ 0} & \makecell{0 \\ 0} & \makecell{0 \\ 0} \\
\rowcolor{gray!40}
\textbf{Total}  & \bfseries\makecell{437 \\ 437} & \bfseries\makecell{1573 \\ 1576} & \bfseries\makecell{781 \\ 786} & \bfseries\makecell{521 \\ 529} & \bfseries\makecell{534 \\ 539} & \bfseries\makecell{273 \\ 276} & \bfseries\makecell{72 \\ 73} &
\bfseries\makecell{5 \\ 5} \\
\hline
\end{tabularx}
\caption{Speaker distribution among different age groups in the training set for female (top) and male (bottom).}
\label{tab:ages:train}
\end{table*}

\begin{table*}[t]
\centering
\footnotesize
\rowcolors{2}{white}{gray!15}
\begin{tabularx}{1.0\textwidth}{b?{1.0pt}s|s|s|s|s|s|s|s}
\toprule
\rowcolor{white}
 & \makecell{Age~\textless~19} & \makecell{19~-~29} & \makecell{30~-~39} & \makecell{40~-~49} & \makecell{50~-~59} & \makecell{60~-~69} & \makecell{70~-~79} & \makecell{80~-~89} \\
\hline
 English  & \makecell{45 \\ 45} & \makecell{135 \\ 144} & \makecell{65 \\ 71} & \makecell{47 \\ 45} & \makecell{47 \\ 49} & \makecell{29 \\ 29} & \makecell{10 \\ 9} & \makecell{0 \\ 1} \\
 German  & \makecell{4 \\ 4} & \makecell{22 \\ 21} & \makecell{16 \\ 17} & \makecell{9 \\ 8} & \makecell{12 \\ 13} & \makecell{5 \\ 5} & \makecell{1 \\ 1} & \makecell{0 \\ 0} \\
 Spanish  & \makecell{12 \\ 12} & \makecell{44 \\ 44} & \makecell{19 \\ 19} & \makecell{13 \\ 13} & \makecell{10 \\ 10} & \makecell{3 \\ 3} & \makecell{0 \\ 0} & \makecell{0 \\ 0} \\
 French  & \makecell{7 \\ 7} & \makecell{31 \\ 31} & \makecell{15 \\ 15} & \makecell{13 \\ 13} & \makecell{13 \\ 13} & \makecell{7 \\ 7} & \makecell{2 \\ 2} & \makecell{0 \\ 0} \\
 Italian  & \makecell{2 \\ 2} & \makecell{14 \\ 14} & \makecell{9 \\ 9} & \makecell{7 \\ 7} & \makecell{7 \\ 7} & \makecell{4 \\ 4} & \makecell{1 \\ 1} & \makecell{0 \\ 0} \\
 Russian  & \makecell{3 \\ 3} & \makecell{8 \\ 8} & \makecell{3 \\ 3} & \makecell{2 \\ 2} & \makecell{1 \\ 1} & \makecell{0 \\ 0} & \makecell{0 \\ 0} & \makecell{0 \\ 0} \\
\rowcolor{gray!40}
\textbf{Total} & \bfseries\makecell{73 \\ 73} & \bfseries\makecell{254 \\ 262} & \bfseries\makecell{127 \\ 134} & \bfseries\makecell{91 \\ 88} & \bfseries\makecell{90 \\ 93} & \bfseries\makecell{48 \\ 48} & \bfseries\makecell{14 \\ 13} & \bfseries\makecell{0 \\ 1} \\
\hline
\end{tabularx}
\caption{Speaker distribution among different age groups in the development set for female (top) and male (bottom).}
\label{tab:ages:dev}
\end{table*}

While CV could be considered a decent corpus for any end-to-end automated speech recognition (ASR) task, there are several important drawbacks. First of all, CV provides only a text transcript as ground truth reference. For training and testing acoustic models, CV does not provide any phonetic transcript or segmentation. Furthermore, the distribution of speakers and speech samples is not ideal in many cases. The English CV 7.0 corpus for example comprised 45\,\% male but only 15\,\% female speakers. For the remaining contributions, gender information was unavailable. Another major problem is the number of samples certain speakers were able to contribute. In the official training split shipped with the previously mentioned English CV dataset, the most overrepresented speaker (according to ID) contributed more than 35,000 samples, which equaled 4.7\,\% of all training samples.\\
This paper introduces \textit{Common Phone} (CP), a refined version of CV, which alleviates the aforementioned drawbacks to provide a corpus that meets modern machine learning (ML) requirements for acoustic modelling in a multi-lingual setup. After a brief summary of the structure of CP, the speaker selection process is explained, including an overview of speaker distributions in the entire dataset. Afterwards, the automated phonetic annotation procedure is described, as well as the utilized phonetic inventory. To validate the quality of the phonetic labels, we fine-tuned a Wav2Vec 2.0 model and tested on CP's test split. The resulting PERs showed that after training with CP, the model was able to reliably predict sequences of phonetic symbols across different languages.

\section{Materials and Methods}

\begin{table*}[t]
\centering
\footnotesize
\rowcolors{2}{white}{gray!15}
\begin{tabularx}{1.0\textwidth}{b?{1.0pt}s|s|s|s|s|s|s|s}
\toprule
\rowcolor{white}
 & \makecell{Age~\textless~19} & \makecell{19~-~29} & \makecell{30~-~39} & \makecell{40~-~49} & \makecell{50~-~59} & \makecell{60~-~69} & \makecell{70~-~79} & \makecell{80~-~89} \\
\hline
  English  & \makecell{43 \\ 47} & \makecell{143 \\ 145} & \makecell{65 \\ 70} & \makecell{44 \\ 46} & \makecell{46 \\ 47} & \makecell{27 \\ 29} & \makecell{10 \\ 10} & \makecell{1 \\ 1} \\
 German  & \makecell{5 \\ 5} & \makecell{20 \\ 19} & \makecell{15 \\ 15} & \makecell{9 \\ 9} & \makecell{13 \\ 11} & \makecell{5 \\ 5} & \makecell{1 \\ 1} & \makecell{1 \\ 1} \\
 Spanish  & \makecell{12 \\ 12} & \makecell{44 \\ 44} & \makecell{19 \\ 19} & \makecell{13 \\ 13} & \makecell{10 \\ 10} & \makecell{3 \\ 3} & \makecell{1 \\ 1} & \makecell{1 \\ 1} \\
 French  & \makecell{7 \\ 7} & \makecell{31 \\ 31} & \makecell{16 \\ 16} & \makecell{13 \\ 13} & \makecell{13 \\ 13} & \makecell{7 \\ 7} & \makecell{2 \\ 2} & \makecell{0 \\ 0} \\
 Italian  & \makecell{2 \\ 2} & \makecell{13 \\ 14} & \makecell{9 \\ 9} & \makecell{8 \\ 8} & \makecell{8 \\ 8} & \makecell{4 \\ 4} & \makecell{1 \\ 1} & \makecell{0 \\ 0} \\
 Russian  & \makecell{4 \\ 4} & \makecell{8 \\ 8} & \makecell{3 \\ 3} & \makecell{2 \\ 2} & \makecell{1 \\ 1} & \makecell{0 \\ 0} & \makecell{0 \\ 0} & \makecell{0 \\ 0} \\
\rowcolor{gray!40}
\textbf{Total} & \bfseries\makecell{73 \\ 77} & \bfseries\makecell{259 \\ 261} & \bfseries\makecell{127 \\ 132} & \bfseries\makecell{89 \\ 91} & \bfseries\makecell{91 \\ 90} & \bfseries\makecell{46 \\ 48} & \bfseries\makecell{15 \\ 15} & \bfseries\makecell{3 \\ 3} \\
\hline
\end{tabularx}
\caption{Speaker distribution among different age groups in the test set for female (top) and male (bottom).}
\label{tab:ages:test}
\end{table*}

\subsection{Corpus structure}
The structure of CP is very similar to that of CV. The directory of each language (English, German, Spanish, French, Italian and Russian) contains CSV-files for the respective train, development and test splits, and an additional one summarizing meta information of all speakers. Directory \textit{mp3} contains the original recordings from CV. These recordings have not been altered in any way. It is important to notice that audio files in CV do not share a common sampling rate (we found 32, 44.1 and 48~kHz), thus varying values should be expected when working with the original recordings.\\
In an additional folder \textit{wav}, raw PCM files were provided through simple decompression of their respective \textit{mp3} counterparts. This was done for two main reasons: Firstly, all files could be converted to a format common in speech signal processing. We chose a sampling rate of 16~kHz, 16 bits depth and mono-channel configuration. Additionally, most existing ML environments and projects expect (or at least support) raw \textit{wav} files as input. As the waveform had been reconstructed from a lossy compression~\cite{brandenburg1999mp3}, it is not to be mistaken for a lossless version of CV recordings.\\
The folder \textit{grids} contains \textit{Praat}~\cite{boersma2001speak} TextGrids with word- and phonetic-level segmentation for every recording.

\subsection{Speaker Selection}
The main objective during speaker selection was to distribute data evenly among languages, genders and age groups, while at the same time keeping as many speakers as possible. Selection was done on the entire set of validated data in CV, omitting the original splits for training, development and test.\\
In a first step, all contributions that did not include information about age and gender had been removed. Not only did this help to keep track of speaker distributions, but it also allowed to assign speakers to only one of the three splits for training, development and test. CV assigns a session ID to every speech donation. If the same speaker donated samples through multiple sessions, it would be impossible to link all contributions to that same speaker. However, if a speaker was logged in to their account on the CV website, all contributions (even over varying sessions) would be linked to a static account ID. Meta information such as age and gender could only be provided through a user account. If that information was available, the recording was made as a logged-in user and the donation could always be linked to a particular contributor.\\
After this pre-selection, all speakers were assigned to a slot within an age-gender grid. From each age slot, contributors were randomly selected in pairs of female and male. The first pair was assigned to the test, the second to the development set, and another five pairs were partitioned into the training split. This procedure was repeated for every age slot until no more speaker pairs were available. After repeating this procedure, speakers were distributed as shown in Tables~\ref{tab:ages:train}, \ref{tab:ages:dev} and ~\ref{tab:ages:test} for training, development and test splits, respectively.\\
In the following step, samples were drawn from each speaker, such that there were as many unique uttered sentences as possible. The number of samples taken from a speaker differed among languages due to their uneven amounts of speakers. We chose to draw 2 (English), 9 (Spanish), 11 (French), 13 (German), 28 (Italian) and 80 (Russian) samples per speaker to ensure that the resulting corpus was not biased towards a particular language. If it was not possible to draw at least one speech sample with a unique text transcript for a particular speaker, that speaker was omitted. For each language, \textit{meta.csv} provides a list of all speakers by their respective ID, which is identical to the one from CV. It summarizes a contributors age group, gender, information about a possible accent if available, and what split the contributor had been assigned to.\\
After sample selection, CP comprised 76,307 speech samples with 73,644 unique texts, totalling 116.5~hours of recorded audio collected from 11,246 unique speakers. The data distribution between languages and splits is summarized in Table~\ref{tab:hours}.

\subsection{Phonetic Inventory}
The entire phonetic inventory used for CP is given in International Phonetic Alphabet~\cite{international1999handbook} (IPA) format and comprises a total of 101 symbols. Table~\ref{tab:ipa} summarizes the core set of symbols, excluding the one for silence and 26 elongated variants. The presented IPA symbols are not to be mistaken for phonemes~\cite{Moore2019}, but rather resemble a set of phones that sufficed to describe the speech of all six languages.\\
Of course, none of the languages required the entire inventory. German (48 symbols) and Italian (47) had the largest inventories, with the former introducing numerous umlauts and the latter differentiating between normal and elongated stops (sustained closure before the burst). Russian (41 symbols) introduced the many palatalized variants of phones. By including French (39 symbols), the entire inventory was enriched by multiple nasalized vowels. English (37) and Spanish (33) were found to be the languages with the smallest inventory.

\begin{table}[t]
\centering
\normalsize
\rowcolors{2}{white}{gray!15}
\begin{tabular}{l|c}
\toprule
 Group & Phonetic Symbols \\
 \hline
 Vowels & \colorbox{gray!15}{\makecell{a\:\~a\:\textturna\:\textscripta\:\textturnscripta\:\ae\:\~\textscoelig\:\textturnv\:e\:\~e\:\textepsilon\:\textrevepsilon\:\textschwa\\i\:\textbari\:\textsci\:o\:\~o\:\o\:\oe\:\textopeno\:\textupsilon\:u\:y\:\textscy}} \\
 Stops & \colorbox{white}{\makecell{b\:b\textsuperscript{j}\:d\:d\textsuperscript{j}\:g\:g\textsuperscript{j}\:p\:p\textsuperscript{j}\:t\:t\textsuperscript{j}\:k\:k\textsuperscript{j}\:\textglotstop}} \\
 Fricatives & \colorbox{gray!15}{\makecell{\textbeta\:\c{c}\:\dh\:f\:f\textsuperscript{j}\:h\:\textctj\:s\:s\textsuperscript{j}\:\textesh\:\textesh\textsuperscript{j}\:v\:v\textsuperscript{j}\\x\:x\textsuperscript{j}\:\textgamma\:z\:z\textsuperscript{j}\:\textyogh\:\texttheta}} \\
 Nasals & \colorbox{white}{\makecell{m\:\textltailm\:m\textsuperscript{j}\:n\:\ng\:\textltailn\:n\textsuperscript{j}}} \\
 Approximants & \colorbox{gray!15}{\makecell{\textturnh\:j\:l\:l\textsuperscript{j}\:w\:\textturny}} \\
 Trills & \colorbox{white}{\makecell{r\:r\textsuperscript{j}\:\textscr}} \\
 \hline
 
\end{tabular}
\caption{Summary of all phonetic IPA symbols used throughout the corpus. Not included in the table but part of the annotation are a symbol for silence as well as 26 elongated variants of presented phones.}
\label{tab:ipa}
\end{table}

\subsection{Phonetic Annotation}
To generate phonetic annotation, we used \textit{WebMAUS}~\cite{kisler2017multilingual}, a web-service provided by the \textit{Bavarian Archive for Speech Signals} (BAS). The Munich AUtomatic Segmentation (MAUS) toolkit provides a routine to reliably predict pronunciation from a pair of speech recording and text transcript~\cite{schiel1999,schiel2015statistical}. The preset pipeline \textit{G2P\_MAUS} without ASR was used with the respective language of the sample, requesting the output phonetic symbols to be encoded in IPA. The pipeline without ASR only disabled the initial ASR for transcription (which was already available), not the one for prediction of pronunciation. After running a grapheme-to-phoneme (G2P) model, MAUS estimated the true pronunciation from the ideal (G2P) and recognized (ASR) pronunciations. The default weight factors for deciding between the two options were left unchanged for all languages.\\
MAUS returned its segmentation result in the form of \textit{Praat} TextGrids. Each contained word- and phonetic-level segmentation of the audio-signal. When choosing IPA as output symbol, phonetic transcription on word-level was given in IPA, but on phonetic level, MAUS yielded X-SAMPA~\cite{Wells1995ComputercodingTI} symbols. As this was just another coding format, translation to IPA was trivial, thus all phonetic symbols in the provided TextGrids follow IPA standards.

\subsection{Acoustic Model Training}
The training split from CP was used to fine-tune a Wav2Vec 2.0~\cite{baevski2020wav2vec} base model. The model had been pre-trained on the 960 hours Librispeech~\cite{panayotov2015librispeech} corpus of read English speech. A final linear layer was added for classification with 102 output nodes, one for each phonetic symbol and an additional one for a blank token to perform connectionist temporal classification (CTC)~\cite{graves2006connectionist}. Adam optimizer~\cite{kingma2014adam} was used with an initial learning rate (LR) of $3\cdot10^{-6}$. During warm-up, the LR increased linearly to $3\cdot10^{-5}$ over the first ten epochs, remained constant for another 30 epochs, and would then decay exponentially by a factor of $0.96$ for the remaining 120 epochs. During a single epoch, the model was shown a subset of 5,000 randomly selected samples from the training set.

\begin{table}[t]
\centering
\rowcolors{2}{white}{gray!15}
\begin{tabular}{l|c|c|c}
\toprule
 Language & Train & Dev & Test \\
 \hline
 English & 14.1 & 2.3 & 2.3 \\
 German & 13.6 & 2.3 & 2.2 \\
 Spanish & 14.5 & 2.5 & 2.6 \\
 French & 14.6 & 2.5 & 2.5 \\
 Italian & 16.5 & 3.0 & 3.1 \\
 Russian & 12.7 & 2.6 & 2.8 \\
 \rowcolor{gray!40}
 \textbf{Total} & \textbf{85.8} & \textbf{15.2} & \textbf{15.5} \\
 \hline
 
\end{tabular}
\caption{Recorded hours of speech in the splits for all six languages.}
\label{tab:hours}
\end{table}

Afterwards, the most probable sequence of phonetic symbols could be estimated through a beam search (beam width~=~10) and CTC decoding. The beam search was not extended with a language model due to the multilingual setup. Despite the fact that our model did not predict phonemes, but phonetic symbols, PER was still considered a decent metric to evaluate the performance of the entire system, as it simply measures the amount of insertions, deletions or replacements required to transform the predicted sequence into the true sequence, relative to the true sequence' length.

\section{Results}
The model predicted phonetic symbols with average PERs of 17.8\,\% on the development and 18.1\,\% on the test set of CP. Results for the different languages are shown in Table~\ref{tab:pers}. English and Spanish were found to be the easiest languages to predict phonetic symbols in our setup. On both development and test, the weakest results were observed for Russian.

\begin{table}[b]
\centering
\rowcolors{2}{white}{gray!15}
\begin{tabular}{l|c|c}
\toprule
 Language & Dev & Test \\
 \hline
 English & 15.5 & 15.6 \\
 German & 19.4 & 19.4 \\
 Spanish & 14.5 & 15.0 \\
 French & 18.8 & 18.4 \\
 Italian & 17.8 & 17.4 \\
 Russian & 20.0 & 21.4 \\
 \hline
 
\end{tabular}
\caption{PERs on development and test for the different languages.}
\label{tab:pers}
\end{table}

Differences between predicted and true sequence had also been analyzed for frequent substitution patterns. The model sometimes struggled to correctly differentiate between short and elongated stop sounds that were common in Italian. For stop phones in general, confusions between voiced and unvoiced productions were also observed more frequently. For Russian, a frequent confusion was those of palatalized phones with their non-palatalized counterparts. In some cases, the model would even append a palatal approximant [/j/ or /\textturny/] after the non-palatalized phone.

\section{Discussion}
There were likely two main reasons for English and Spanish to yield the lowest PERs. Firstly, the entire Wav2Vec model was pre-trained on almost a thousand hours of English speech. This could potentially induce some bias towards English pronunciation. For Spanish, the fact that the language used comparatively few stop phones (12.8\,\% compared to the other languages [18.3\,\%~--~21.4\,\%] as estimated from the ground truth) Finally, Spanish and English comprised the smallest phonetic inventories with 33 and 37 symbols, respectively.\\
The weaker result for Russian did not come as a surprise, because the language introduced a large number of new phones to the inventory that were not used among the other datasets.\\
A PER of 18.1\,\% may appear a lot higher than the one reported for Wav2Vec 2.0 mentioned earlier (8.3\,\%). This was, however, calculated on TIMIT, which includes recordings from only one microphone in a single language, collected in a clean acoustic environment. Furthermore, when computing PERs on TIMIT, it is common to collapse the annotated phonemes to 39 classes. This was done for the Wav2Vec 2.0 evaluation as well. The presented model trained with CP had to distinguish between 101 classes from six different languages. Lastly, the single-digit PER was achieved only by the large version of Wav2Vec, which comprised 317 million parameters. For this study, the base model with 95 million parameters was used, as we were not interested in showing peak PER values, but intended to validate the suitability of phonetic labels created with \textit{WebMAUS}.\\
Finally, some light should be shed on the potential weaknesses of CP. During the process of speaker selection, the session ID was used as the unique identifier for a speaker. While we managed to ensure that this ID was constant throughout multiple sessions by only including users who had been logged in to their account, we could not rule out the possibility of multiple speakers contributing through the same account. However, no such case was identified during manual investigation.\\
It could also happen that the predicted pronunciation from \textit{WebMaus} was not entirely correct, or that a file contained longer segments of silence, background noise, or poorly intelligible speech which confused the ASR during the alignment. Whilst these artifacts could result in a certain amount of label noise (which can be found in almost every ML dataset), they would also allow a model to explore a much larger space of acoustic conditions, speaker traits and noise patterns, ultimately resulting in a more robust decision function.

\section{Conclusion}
CP provides a reduced, gender-balanced version of CV comprising six different languages. All samples received phonetic annotation, which could be proven to yield reliable results when used for training a state-of-the-art acoustic model. Totalling over 100~h of recorded speech collected from more than 11,000 contributors in unsupervised recording conditions, CP can help to bridge the gap between research and application of acoustic models.\\
The corpus is distributed free of copyright under a \textit{Creative Commons (CC0 1.0 Universal)} license, just like CV itself. It is distributed via \nolinkurl{www.zenodo.org} (doi:~10.5281/zenodo.5846137). In the future, the phonetic inventory should grow further by adding more languages. CP should also receive updates along with major revisions of CV.

\section{Bibliographical References}\label{reference}

\bibliographystyle{lrec2022-bib}
\bibliography{lrec2022-example}

\section{Language Resource References}
\bibliographystylelanguageresource{lrec2022-bib}
\bibliographylanguageresource{languageresource}

\end{document}